\documentstyle[12pt,epsfig,cite]{article}

\newcommand{\ptmiss}{{\mbox{$\not\hspace{-.55ex}{P}_T$}}}
\newcommand{\squark}{\mbox{${\tilde q}$}}

\newcommand{\rpvio}{{\not\hspace{-.55ex}{R}_P}}

\begin{document}

\title {\bf\LARGE Search for Resonances Decaying to $e^+$-jet
in $e^+p$ Interactions at HERA }

\author{ZEUS Collaboration}

\date{}

\maketitle

\begin{abstract}
\noindent The $e^+$-jet invariant 
mass spectrum produced in the reaction $e^+ p \rightarrow e^+ X$ has been
studied at a center-of-mass energy of 300~GeV.
The data were
collected using the ZEUS detector operating at the HERA collider, and 
correspond to an integrated luminosity of $47.7$~pb$^{-1}$.
The observed mass spectrum is in good agreement with Standard Model 
expectations up to an $e^+$-jet mass of $210$~GeV.  Above this mass,
some excess is seen.  The angular distribution of these events is
 typical of high-$Q^2$ neutral current events and does not give
convincing evidence for the presence of a narrow scalar or vector 
state.
Limits are presented on the product of cross section and branching ratio
for such a state and are interpreted
as limits on leptoquark or R-parity-violating squark production.
Specific leptoquark types are ruled out at $95$\% confidence level
for coupling strength $\lambda=0.3$
for masses between $150$ and  $280$~GeV.

\end{abstract}

\vspace{-14.5cm}
\begin{flushleft}
\tt DESY 00-023 \\
February 2000 \\
\end{flushleft}

{\hspace*{-0mm} } \pagestyle{plain} \thispagestyle{empty} \newpage 

\topmargin-1.cm                                                                                    
\evensidemargin-0.3cm                                                                              
\oddsidemargin-0.3cm                                                                               
\textwidth 16.cm                                                                                   
\textheight 680pt                                                                                  
\parindent0.cm                                                                                     
\parskip0.3cm plus0.05cm minus0.05cm                                                               
\def\3{\ss}                                                                                        
\newcommand{\address}{ }                                                                           
\pagenumbering{Roman}                                                                              

\begin{center}                                                                                     
{                      \Large  The ZEUS Collaboration              }                               
\end{center}

%
%
%
%

  J.~Breitweg,                                                                                     
  S.~Chekanov,                                                                                     
  M.~Derrick,                                                                                      
  D.~Krakauer,                                                                                     
  S.~Magill,                                                                                       
  B.~Musgrave,                                                                                     
  A.~Pellegrino,                                                                                   
  J.~Repond,                                                                                       
  R.~Stanek,                                                                                       
  R.~Yoshida\\                                                                                     
 {\it Argonne National Laboratory, Argonne, IL, USA}~$^{p}$
% ...
\newpage                                                                                           
$^{\    1}$ now visiting scientist at DESY \\                                                      
% ...
                                                           %
                                                           %
\newpage   
                                                           %
                                                           %
\begin{tabular}[h]{rp{14cm}}                                                                       
$^{a}$ &  supported by the Natural Sciences and Engineering Research                               
          Council of Canada (NSERC)  \\                                                            
$^{b}$ &  supported by the FCAR of Qu\'ebec, Canada  \\                                            
$^{c}$ &  supported by the German Federal Ministry for Education and                               
          Science, Research and Technology (BMBF), under contract                                  
          numbers 057BN19P, 057FR19P, 057HH19P, 057HH29P, 057SI75I \\                              
$^{d}$ &  supported by the MINERVA Gesellschaft f\"ur Forschung GmbH, the                          
German Israeli Foundation, and by the Israel Ministry of Science \\                                
$^{e}$ &  supported by the German-Israeli Foundation, the Israel Science                           
          Foundation, the U.S.-Israel Binational Science Foundation, and by                        
          the Israel Ministry of Science \\                                                        
$^{f}$ &  supported by the Italian National Institute for Nuclear Physics                          
          (INFN) \\                                                                                
$^{g}$ &  supported by the Japanese Ministry of Education, Science and                             
          Culture (the Monbusho) and its grants for Scientific Research \\                         
$^{h}$ &  supported by the Korean Ministry of Education and Korea Science                          
          and Engineering Foundation  \\                                                           
$^{i}$ &  supported by the Netherlands Foundation for Research on                                  
          Matter (FOM) \\                                                                          
$^{j}$ &  supported by the Polish State Committee for Scientific Research,                         
          grant No. 112/E-356/SPUB/DESY/P03/DZ 3/99, 620/E-77/SPUB/DESY/P-03/                      
          DZ 1/99, 2P03B03216, 2P03B04616, 2P03B03517, and by the German                           
          Federal Ministry of Education and Science, Research and Technology (BMBF)\\              
$^{k}$ &  supported by the Polish State Committee for Scientific                                   
          Research (grant No. 2P03B08614 and 2P03B06116) \\                                        
$^{l}$ &  partially supported by the German Federal Ministry for                                   
          Education and Science, Research and Technology (BMBF)  \\                                
$^{m}$ &  supported by the Fund for Fundamental Research of Russian Ministry                       
          for Science and Edu\-cation and by the German Federal Ministry for                       
          Education and Science, Research and Technology (BMBF) \\                                 
$^{n}$ &  supported by the Spanish Ministry of Education                                           
          and Science through funds provided by CICYT \\                                           
$^{o}$ &  supported by the Particle Physics and                                                    
          Astronomy Research Council \\                                                            
$^{p}$ &  supported by the US Department of Energy \\                                              
$^{q}$ &  supported by the US National Science Foundation                                          
\end{tabular}                                                                                      
                                                           %




\newpage

\newpage

\topmargin-1.5cm
\evensidemargin-0.3cm
\oddsidemargin-0.3cm
\textwidth 16.cm
\textheight 650pt
\parindent0.cm
\parskip0.3cm plus0.05cm minus0.05cm

\pagenumbering{arabic} 
\setcounter{page}{1}

\section{Introduction}

The $e^+$-jet mass spectrum in $e^+p$ scattering has been investigated
with the ZEUS detector at HERA.  An excess of events relative to
Standard Model expectations has previously been reported by the 
H1~\cite{ref:H1highx,ref:H1_limits}
and ZEUS~\cite{ref:ZEUShighx,ref:highQ2}
collaborations in neutral current
deep inelastic scattering at high $x$ 
and high $Q^2$.  These events contain high-mass $e^+$-jet final states.
Several models have been
discussed~\cite{ref:dokshitzer} as possible sources of these events, 
including leptoquark production~\cite{ref:lq} and R-parity-violating squark 
production~\cite{ref:Rpsusy}.
This paper presents an analysis of ZEUS data specifically aimed at 
searching for high mass states decaying to $e^+$-jet.

Candidate events with high transverse energy, an
identified final-state positron, and at least one jet are selected. 
The measured energies
($E_e',E_j$) and angles of the final-state positron 
and the jet with highest transverse momentum are used to calculate an
invariant mass 
\begin{equation}
M_{ej}^{2}=2E_{e}^{\prime }E_j\cdot (1-\cos\xi )\;,  \label{eq:Mass}
\end{equation}
where $\xi$ is the angle between the positron and jet.
The angle between the outgoing and incoming
positron in the $e^+$-jet rest frame, $\theta^*$, is also reconstructed
using the measured energies and angles.  No
 assumptions about the production process are made in the reconstruction
of either $M_{ej}$ or $\theta^*$.

The search was performed using $47.7$~pb$^{-1}$ of data collected in the
1994-1997 running periods.  In the following, expectations from
the Standard Model,  leptoquark production 
and  R-parity-violating squark
production are summarized.  After a discussion of the
experimental conditions, the analysis is described and the 
$M_{ej}$ and $\cos \theta^{\ast}$ distributions presented.  Since these 
distributions do not show a clear signal for a narrow resonance,
limits on the cross section times branching ratio are extracted for the 
production of such a state.  Limits are also presented in the mass versus 
coupling plane which can be applied to leptoquark and squark production.

\section{Model Expectations}
\begin{figure}[hbpt]
\begin{center}
\epsfig{figure=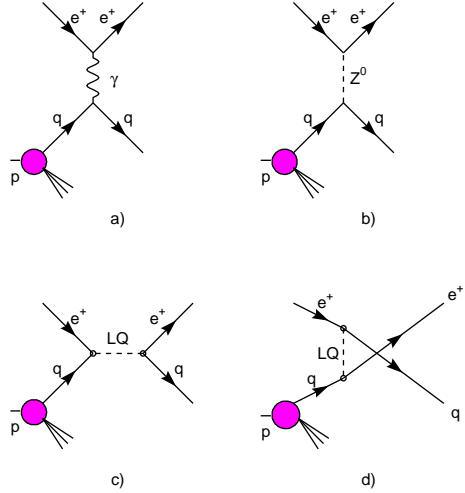,height=8.cm}
\end{center}
\caption{\it Diagrams for NC scattering via a) photon exchange and 
b) $Z^0$ exchange. The leptoquark diagrams for the same initial and final
states are c) s-channel LQ 
production for fermion number
$F=0$ LQ and d) u-channel LQ exchange for an $F=2$ LQ. }
\label{fig:NCLQ}
\end{figure}

High-mass $e^+$-jet pairs, produced in the Standard Model (SM)
via neutral current (NC) scattering, form the principal background
to the search for heavy states. This process is reviewed first.
Leptoquark (LQ) production
and squark production in R-parity-violating ($\rpvio$) supersymmetry
are used as examples of physics beyond the SM that could generate 
the $e^+$-jet final state. The diagrams
for NC and LQ processes are shown in Fig.~\ref{fig:NCLQ}. 
The squark
production diagrams are similar to the LQ diagrams, but different
decay modes are possible, as discussed below.

\subsection{Standard Model Expectations}
The kinematic variables used to describe the deep inelastic 
scattering (DIS) reaction
\begin{displaymath}
e^+p \rightarrow e^+X
\end{displaymath}
are
\begin{eqnarray}
Q^2 & = & -q^2 = -(k-k')^2 \;, \\
y & = & \frac{q\cdot P}{k\cdot P} \hskip 3.cm {\rm and} \\
x & = & \frac{Q^2}{2 q\cdot P} \; ,
\end{eqnarray}
where $k$ and $k'$ are the four-momenta of the incoming and outgoing positron,
respectively, and $P$ is the four-momentum of the incoming proton.  The
center-of-mass energy is given by $s=(k+P)^2 \approx (300~{\rm GeV})^2$.
The NC interaction occurs between the positron and a parton
(quark) inside the proton (see Fig.~\ref{fig:NCLQ}).
The production of the
large $e^+$-jet masses of interest requires high $x$ partons,
where the valence quarks dominate the proton structure.

In leading-order electroweak theory, 
the cross section for the NC DIS reaction can be expressed 
as~\cite{ref:SF}
\begin{equation}
\label{eq:DIS}
\frac{d^2 \sigma(e^+p)}{dxdy} = \frac{2\pi \alpha^2}{sx^2y^2}
\left[Y_+ {\it F_2} - Y_-x{\it F_3} + y^2F_L \right]
\end{equation}
with $Y_{\pm} = 1\pm(1-y)^2$ and $\alpha$ the fine structure constant. 
The contribution from
the longitudinal structure function,
$F_L$, is expected to be negligible in the
kinematic range considered here.

The
$x$ dependence of the NC cross section is very steep.  In addition to the
explicit $1/x^2$ factor, the structure functions $F_2$ and $xF_3$
are dominated at large $x$ 
by valence-quark densities 
that fall quickly for $x >0.3$.
The $y$ dependence of the cross
section is dominated by the $1/y^2$ term.  The structure functions vary
slowly with $y$ at fixed $x$.
The uncertainty in the NC cross section predicted by Eq.~\ref{eq:DIS}
is dominated by the uncertainty in the structure functions (parton densities),
and is small, about $5$\% at the
high-$x$ and moderate-$y$ ranges of this analysis~\cite{ref:highQ2}.
The quantity of interest in this paper is the $e^+$-jet cross section,
which is sensitive to QCD corrections. The
uncertainty arising from these corrections has been estimated to be small for this analysis.

For DIS or LQ events produced via
the diagrams shown in Fig.~~\ref{fig:NCLQ} (i.e.
assuming no QED or QCD radiation), the mass of the $eq$ system is
related to $x$ via

\begin{equation}
M^2 = sx
\end{equation} 
and $ \theta^{\ast}$ is related to $y$ via

\begin{equation}
\label{eq:y_cost}
\cos \theta ^{\ast }=1 - 2y  \;.
\end{equation}

The steeply-falling  $x$ and $y$ dependences of DIS events will therefore 
produce distributions falling sharply with mass and peaking towards
$\cos \theta ^{\ast } = 1$.

\subsection{Leptoquark Production and Exchange}
\label{sec:LQ}

Leptoquark production is an example of new physics that could
generate high-mass $e^+$-jet pairs. 
The set of leptoquarks with $SU(3)\times SU(2)\times U(1)$-invariant
couplings has been specified \cite{ref:lq}.
Only LQs with fermion number $F=L+3B=0$ are considered here, 
where $L$ and $B$ denote the lepton
and baryon number, respectively.
These leptoquarks
are listed in Table~\ref{tab:lq} together with some of their properties.
The $F=0$ LQs have higher cross sections in $e^{+}p$
scattering than $e^-p$ scattering
since in the $e^+p$ case a valence quark can fuse with the positron.

\begin{table}[h]
\begin{displaymath}
\begin{tabular}{c|cccc}

LQ species & $q$ & Production & Decay & Branching ratio \\ \hline
$S_{1/2}^L$ & -5/3 & $e_L\bar{u}$ & $e\bar{u}$ & 1  \\ 
$S_{1/2}^R$ & -5/3 & $e_R\bar{u} $ & $e \bar{u}$ & 1  \\ 
& -2/3 & $e_R\bar{d} $ & $e\bar{d}$ & 1  \\ 
${\tilde S}_{1/2}^L$ & -2/3 & $e_L\bar{d} $ & $e\bar{d}$ & 1  \\ 
$V_0^L$ & -2/3 & $e_L\bar{d} $ & $e\bar{d}$ & 1/2  \\ 
&  & 
$$
& $\nu_e\bar{u}$ & 1/2 \\ 
$V_0^R$ & -2/3 & $e_R\bar{d} $ & $e\bar{d}$ & 1 \\ 
${\tilde V}_0^R$ & -5/3 & $e_R\bar{u} $ & $e\bar{u}$ & 1 \\ 
$V_1^L$ & -5/3 & $e_L\bar{u} $ & $e\bar{u}$ & 1 \\ 
& -2/3 & $e_L\bar{d} $ & $e\bar{d}$ & 1/2 \\ 
&  & 
$$
& $\nu_e\bar{u}$ & 1/2 \\ \hline
\end{tabular}
\end{displaymath}
\caption{\it The $F=0$ leptoquarks that can be produced at HERA.
The LQ species are divided according to their spin ($S$
for scalar and $V$ for vector), their chirality ($L$ or $R$) and their weak
isospin ($0,1/2,1$). The leptoquarks ${\tilde S}$ and ${\tilde V}$ differ by
two units of hypercharge from $S$ and $V$, respectively. In addition, the
electric charge, $q$, of the leptoquarks, the production channel, as well as
their allowed decay channels assuming lepton-flavor conservation,
are displayed. The quantum numbers and decay channels correspond to an
electron-type LQ. For positrons, the corresponding anti-leptoquarks
have the sign of the electric charge reversed, the helicity of the 
incoming lepton reversed and antiquarks are
replaced by the corresponding quark.
The nomenclature follows the Aachen convention~\protect\cite{ref:LG}.}
\label{tab:lq}
\end{table}

In principle, additional LQ types can be defined~\cite{ref:Davidson}\ which
depend on the generations of the quarks and leptons to which they couple. 
Only LQs which preserve lepton flavor and which couple to first-generation
quarks are considered in this analysis.

As shown in Fig.~\ref{fig:NCLQ}, leptoquark production can generate an 
s-channel resonance provided $m_{LQ}<\sqrt{s}$.
Contributions to the $e^+p$ cross section would also result from
u-channel exchange and
interference of LQ diagrams with photon and $Z^0$ exchange.
The cross section in the presence of a leptoquark can be written as

\begin{equation}
\label{eq:terms}
\frac{d^2 \sigma(e^+p)}{dxdy}  = 
\frac{d^2\sigma^{NC}}{dxdy} + 
\frac{d^2\sigma^{Int}_{u/NC}}{dxdy} + 
\frac{d^2\sigma^{Int}_{s/NC}}{dxdy} + 
\frac{d^2\sigma^{LQ}_{u}}{dxdy} +
\frac{d^2\sigma^{LQ}_{s}}{dxdy} \; .
\end{equation} 

The first term on the right-hand side of Eq.~\ref{eq:terms}
represents the SM contribution discussed previously.  
The second (third) term arises from the
interference between the SM and u-channel (s-channel)
LQ diagram, and the fourth (fifth) term represents
the u-channel (s-channel) LQ diagram alone.  The additional 
contributions to the SM cross section 
depend on two parameters: $m_{LQ}$, and
$\lambda_{R}$ or $\lambda_{L}$, the coupling to 
$e_{L,R}^+$ and quark. Leptoquarks
of well-defined helicity ($\lambda_{R}\cdot \lambda_{L}=0$)
are assumed for simplicity in the limit-setting procedure,
and one species of LQ is assumed to dominate the cross section.
The $\cos \theta^{\ast}$ dependence varies strongly for the
different terms: it is flat for scalar-LQ production in the s-channel,
and for vector-LQ exchange in the u-channel, while it varies as
$(1+ \cos \theta^{\ast})^2$ for vector-LQ production in the
s-channel or scalar-LQ exchange in the u-channel.  The interference
terms produce a $\cos \theta^{\ast}$ dependence which is steeper due
to the sharply-peaking $\cos \theta^{\ast}$ distribution in NC DIS.

In general, the s-channel term dominates the additional contributions to the
SM cross section 
if $m_{LQ}<\sqrt{s}$, the coupling $\lambda$ is small, and the
LQ is produced from a quark rather than an antiquark.  However, there
are conditions for which the other terms can become significant, or even
dominant~\cite{ref:DURHAM},  
leading to important  consequences for
the expected mass spectra and decay angular distributions.
The u-channel and interference terms cannot produce a resonance
peak  in the
mass spectrum and the angular distributions from such terms can behave 
more like those
of NC deep inelastic scattering.  Limits are presented in this paper
for narrow-width LQ and under conditions for which the s-channel term 
dominates.

The width of a LQ depends on its spin and decay modes,
and is proportional to $m_{LQ}$ times the square of the coupling.
In the narrow-width approximation, the
LQ production cross section is given by integrating the s-channel 
term~\cite{ref:lq}:
\begin{equation}
\label{eq:NWA}
\sigma^{NWA} = 
(J+1)\frac{\pi}{4s} \lambda^2 q(x_0,\mu)   
\end{equation}
where $J$ represents the spin of the LQ,
$q(x_0,\mu)$ is the quark density evaluated at
$x_0=m_{LQ}^2/s$ and with the scale $\mu=m_{LQ}^2$.  
In the limit-setting procedure (Sect.~\ref{sec:limits}),
this cross section was corrected for expected QED 
and QCD (for scalar LQ only)
radiative effects.
The QCD corrections~\cite{ref:lqQCD} 
enhance the cross section by 20 - 30\% for the $F=0$ LQ
considered here.  The effect of QED radiation
on the LQ production cross section was calculated
and was found to decrease the
cross section by 5-25\% as $m_{LQ}$ increases 
from $100 \rightarrow 290$~GeV.

\subsection{R-Parity-Violating Squark Production}
\label{sec:Rpvio}

In the supersymmetry (SUSY) superpotential, $R$-parity-violating terms of
the form \linebreak
$\lambda'_{ijk}L^i_LQ^j_L{\overline
D}^k_R$~\cite{ref:Rpsusy} are of particular interest 
for lepton-hadron collisions. 
Here, $L_L$, $Q_L$, and
${\overline D}_R$ denote left-handed lepton and quark doublets and the
right-handed down-type quark-singlet chiral superfields, respectively. The
indices $i$, $j$, and $k$ label their respective
generations.

For $i=1$, which is the case for $ep$ collisions, these operators
can lead to
${\tilde u}$- and ${\tilde d}$-type squark production.  
There are 9 possible production couplings probed in $e^+p$
collisions, corresponding to the reactions~\cite{ref:Hewett_SUSY}
\begin{eqnarray}
e^+ + \bar{u}_j & \rightarrow & \tilde{\bar{d}}_k \;\; , \\
e^+ + d_k & \rightarrow & \tilde{u}_j  \;\; .
\end{eqnarray}
\noindent For production and decay via the $\lambda'_{1jk}$ coupling,
squarks behave like scalar leptoquarks and the final state is
indistinguishable, event by event, from Standard Model neutral and
charged current events.  However, as for the scalar leptoquarks, the
angular distributions of the final-state lepton and quark will be
different and this fact can be exploited in performing searches.  Limits
derived for scalar LQ production can then be directly related to limits
on squark production and decay via $\lambda'_{ijk}$.
In addition to the Yukawa couplings, gauge couplings also exist
whereby $\squark$ can decay by radiating a neutralino or chargino
which can subsequently decay.  The final-state signature depends
on the properties of the neutralino or
chargino.  The search for such decay topologies from a squark is 
outside the scope of this analysis.

\section{Experimental Conditions}
\label{sec:conditions}

In 1994-97, HERA operated with protons of energy
$E_p = 820$~GeV and positrons of energy $E_e = 27.5$~GeV.  The
ZEUS detector is described in detail in~\cite{ref:ZEUS}.
The main components used in the present analysis were the central
tracking detector (CTD) positioned in a 1.43~T solenoidal magnetic field
and the uranium-scintillator sampling calorimeter (CAL).  The
CTD was used to establish an interaction vertex
with a typical resolution along (transverse to) the beam direction
of $0.4 \; (0.1)$~cm.  It was also used in the positron-finding algorithm
that associated a charged track with an energy deposit in the 
calorimeter. The CAL was used to measure the
positron and hadronic energies. The CAL
consists of a forward part (FCAL), a barrel part (BCAL) and a rear part  
(RCAL), with depths of $7,\; 5\; {\rm and} \; 
4$ interaction lengths, 
respectively. The FCAL and BCAL are segmented longitudinally 
into an electromagnetic section (EMC), and two hadronic sections  (HAC1,2).  
The RCAL has one EMC and one HAC section.  The cell structure is formed by  
scintillator tiles;  cell sizes range from $5 \times 20$~cm$^2$  (FEMC) to 
$24.4 \times 35.2$~cm$^2$ at the front face of a BCAL HAC2 cell. The light 
generated in the scintillator is collected on both sides of the module by 
wavelength-shifter bars, allowing a coordinate measurement based on 
knowledge of the attenuation length in the scintillator.  The light is 
converted into an electronic signal by photomultiplier tubes.
The cells are arranged into towers consisting of $4$ EMC cells, a HAC1
cell and a HAC2 cell (in FCAL and BCAL).  The transverse dimensions
of the towers in FCAL 
are $20\times 20$~cm$^2$.  One tower is absent at the center of the FCAL
and RCAL to allow space for passage of the beams.  The outer boundary of the
inner ring of FCAL towers, used to define a fiducial cut for the jet
reconstruction, defines a box of $60\times 60$~cm$^2$.

Under test beam conditions, the CAL has energy resolutions of 
$\sigma/E=0.18/\sqrt{E}$ for positrons hitting the center of a calorimeter
cell and
$\sigma/E=0.35/\sqrt{E}$ for single hadrons, where energies are in GeV.  
In the ZEUS detector,
the energy measurement is affected by the energy loss in the 
material between the interaction point and the calorimeter.
For the events selected
in this analysis, the positrons predominantly strike the 
BCAL, while the jets hit the FCAL.  The in-situ positron-energy 
resolution in
the BCAL has been determined to average 
$\sigma/E = 0.32/\sqrt{E} \oplus 0.03$
while the jet-energy resolution in the FCAL averages
$\sigma/E = 0.55/\sqrt{E}\oplus 0.02$. The jet-energy resolution was
determined by comparing reconstructed jet energies in the calorimeter
with the total energy
of the particles in the hadronic final-state using Monte 
Carlo simulation, and therefore includes small
contributions from the jet-finding algorithm.

In the reconstruction of the positron and jet energies, 
corrections were applied for inactive materials located in front of the
calorimeter and for non-uniformities in the calorimeter response 
\cite{ref:highQ2}.  For the high energies important in this analysis,
the overall energy scale is known to $1$\% for
positrons in BCAL and $2$\% for hadrons in FCAL and BCAL.
The electromagnetic energy
scale was determined by a comparison with 
momentum measurements in the central tracking
detector (using lower-energy electrons and positrons).  Its linearity
was checked with energies reconstructed from
the double angle (DA) method~\cite{ref:DA}.
The hadronic-energy scales in the FCAL and BCAL were determined by
using transverse-momentum balance in NC DIS events.

The angular reconstruction was performed using a combination of 
tracking and calorimeter information.  From Monte Carlo
studies, the polar-angle resolutions were found to be 
$2.5$~mrad for positrons and approximately 
($220/\sqrt{E}-4$)~mrad
for jets with energies above $100$~GeV.

The luminosity was measured from the rate of the bremsstrahlung
process $e^+ p \rightarrow e^+ p \gamma$~\cite{ref:lumi}, 
and has an uncertainty of $1.6$\%.

The ZEUS coordinate system is right-handed and centered on the nominal 
interaction point, with the $Z$ axis pointing in the
direction of the proton beam (forward) and the $X$ axis pointing horizontally
toward the center of HERA. The polar angle $\theta$ is defined with respect
to the $Z$ axis.

\section{Event Selection}

The events of interest with large $e^+$-jet mass
 contain a final-state positron
at a large angle and of much higher energy than that of the incident
positron beam, as well as one or more energetic jets. The only important
SM source of such events is NC
scattering with large $Q^{2}$.  Other potential
backgrounds, such as high transverse-energy ($E_T$) 
photoproduction, were determined to be
negligible.

The following requirements selected events of the
desired topology:

\begin{itemize}
\item A reconstructed event vertex was required in the
range $|Z|<50$~cm.

\item  The total transverse energy, $E_{T}$, was
required to be at least $60$~GeV.

\item  An identified~\cite{ref:highQ2} positron was required
with energy $E_{e}^{\prime }>25$~GeV, located
either in the FCAL or BCAL.  The positron
was required to be well-contained in the BCAL or FCAL and not to point to 
the BCAL/FCAL interface, at approximately 
$31^{\circ} <\theta <36^{\circ}$.
Positrons within $1.5$~cm of the boundary between adjacent BCAL modules,
as determined by tracking information, were also discarded to 
remove showers developing in the wavelength-shifter bars.

\item  A hadronic jet with transverse momentum 
$P^{j}_{T}>15$~GeV, located in a region of good containment, was
required. The jets were reconstructed using the longitudinally-invariant 
$k_T$-clustering
algorithm~\cite{ref:kT1} in the inclusive mode~\cite{ref:kT2}.
Only jets with a reconstructed centroid 
outside the inner ring of FCAL towers
were considered. In events where multiple jets were
reconstructed, the jet with highest transverse momentum was used. 
After all cuts, 12\%  of the events had more than one jet,
both in the data and Monte Carlo simulation (see below).
\end{itemize}

\begin{figure}[hbpt]
\begin{center}
\epsfig{figure=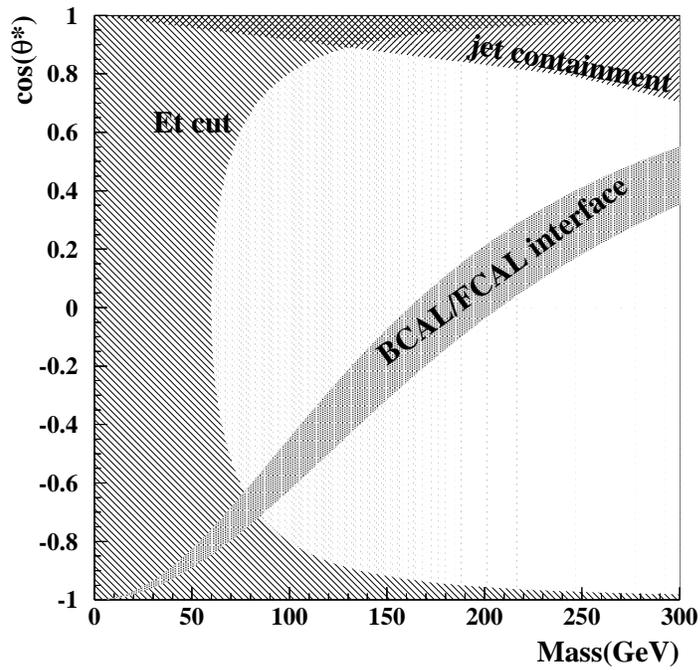,height=10.cm}
\end{center}
\caption{\it The acceptance region (unshaded) in the 
$\cos \theta^{\ast}$ versus $M_{ej}$ plane
allowed by the $E_T$, jet-containment and
positron-fiducial-volume cuts, assuming $eq \rightarrow eq$
scattering at the nominal interaction point.  No detector
simulation is included.}
\label{fig:Acceptance}
\end{figure}

The $E_T$ cut, the jet-containment cut and the positron-containment
cut define the available kinematic region for further analysis, as shown
in Fig.~\ref{fig:Acceptance}.  The jet containment cut, in particular,
limits the values of $\cos\theta ^{\ast }$ that can be measured at
the highest $e^+$-jet masses.  Because most such events have   
$\cos\theta ^{\ast }$ near $1$,
the acceptance for NC DIS events (with
$E_T>60$~GeV) falls below
$10$\% for masses beyond 220~GeV.
 In the region allowed by the cuts shown in Fig.~\ref{fig:Acceptance},
the acceptance is typically $80$\%.

\begin{figure}[hbp]
\begin{center}
\epsfig{figure=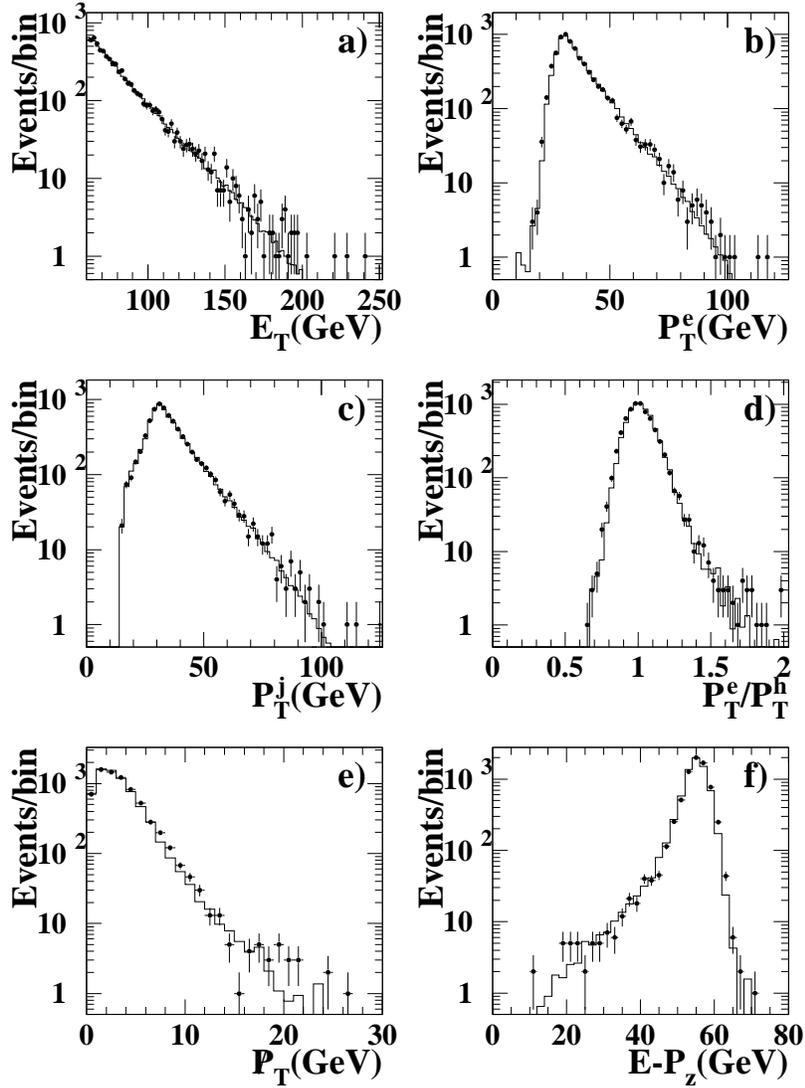,height=16.cm}
\end{center}
\caption{\it Comparison of data (points)
with Standard Model expectations (histograms) for
selected distributions: a) total transverse energy, $E_{T}$; 
b) positron transverse momentum, $P_{T}^{e}$;
 c) jet transverse momentum, $P_{T}^{j}$%
; d) the ratio of the positron to hadron transverse momenta, 
$P_{T}^{e}/P_{T}^{h}$, (e) the missing transverse momentum, \ptmiss, 
and (f) the longitudinal-momentum variable, $E-P_{Z}$, for the event.}
\label{fig:compare}
\end{figure}

A total of $7103$ events remained after applying all cuts, compared to
$6949\pm 445$ events predicted by the NC Monte Carlo simulation 
based on the measured luminosity of $47.7$~pb$^{-1}$
(the sources of uncertainty on the expected number of events are
described in Sect.~\ref{sec:systematics}). \ 
The $E_T$ distributions for data and NC
simulation are compared 
in Fig.~\ref{fig:compare}a. The positron transverse-momentum 
($P_{T}^{e}$) 
spectrum, jet transverse-momentum ($P_{T}^{j}$) spectrum 
and the ratio $P_{T}^{e}/P_{T}^{h}$, where $P_{T}^{h}$ is the transverse 
momentum of the hadronic system, are shown in Figs.~\ref
{fig:compare}(b-d), respectively. The missing transverse 
momentum, \ptmiss, and the longitudinal momentum variable, $E-P_{Z}$ , 
are compared in Figs.~\ref{fig:compare}(e,f). The global properties
of the events are well reproduced by the simulation.

\section{Event Simulation}
\label{sec:MC}

The SM deep inelastic scattering events
were simulated using the HERACLES 4.5.2~\cite{ref:HERACLES} program
with the DJANGO 6 version 2.4~\cite{ref:DJANGO} interface to the 
hadronization
programs.  In HERACLES, corrections for initial- and final-state 
electroweak radiation,
vertex and propagator corrections, and two-boson exchange are included.
The NC DIS hadronic final state was simulated using the MEPS model of
LEPTO 6.5~\cite{ref:MEPS}, which includes order $\alpha_S$
matrix elements and models
of higher-order QCD radiation. As a systematic check, 
the NC final state was simulated using the
color-dipole model of ARIADNE 4.08~\cite{ref:ARIADNE}.
The CTEQ4 parton-distribution set~\cite{ref:CTEQ4} 
was used to evaluate the expected number of events from NC DIS scattering.

The leptoquark events were generated using PYTHIA 6.1~\cite{ref:PYTHIA}.
This program takes into account the finite width of the LQ, but only
includes the s-channel diagrams.  Initial-
and final-state QCD radiation from the quark
and the effect of LQ hadronization before decay
are taken into account, as are initial- and final-state
QED radiation from the positron.

The generated events were input into a GEANT-based~\cite{ref:GEANT}
program which simulated the response of the ZEUS detector.  
The trigger and
offline processing requirements applied to the data were applied to the 
simulated events.  The luminosity of the NC Monte Carlo samples ranges
from $46$~pb$^{-1}$ at $Q^2=400$~GeV$^2$ to $7.3\cdot 10^6$~pb$^{-1}$
at $Q^2=50000$~GeV$^2$.

\section{Mass and $\theta ^{\ast }$ Reconstruction}
\label{sec:Reconstruction}

The mass of each $e^+$-jet pair was reconstructed from
the measured energies and angles of the positron and jet as
described by Eq.~\ref{eq:Mass}.
This formula makes no correction for the finite jet mass. 
Possible mass shifts and the resolutions for resonant lepton-hadron states 
were estimated from PYTHIA.
Narrow scalar LQ events in the mass range $150-290$~GeV were simulated. 
The mean mass for reconstructed events was found to be within 6\% of
the generated value, while the peak position as determined by a Gaussian fit was
typically lower than the generated value by only 1\%. 
The average mass resolution,
determined from a Gaussian fit to the peak of the reconstructed mass
spectrum, ranged from 5.5\% to 3\% for masses from 150 to 290~GeV. The RMS
of the distribution was typically twice as large.  

The positron scattering angle in the $e^+$-jet rest frame, $\theta ^{\ast }$,
was reconstructed as the angle between the incoming and outgoing positron
directions in this frame.  These directions were determined by performing a 
Lorentz transformation using
the measured positron and jet energies and angles in the laboratory
frame. The resolution
in $\cos \theta ^{\ast }$ near $|\cos \theta ^{\ast }|=1$,
as determined from a Gaussian fit, was $0.01$ degrading to $0.03$ as 
$|\cos \theta ^{\ast }|$ decreases.  The
shift in $\cos \theta ^{\ast }$ was less than $0.01$ for both the NC MC and the
leptoquark MC.

In order to determine limits on leptoquark and squark production, the
mass of the electron-hadron system was reconstructed by the constrained-mass method.
This method reconstructs the 
$e^+$-hadron mass as
\begin{eqnarray}
M_{CJ}  & = & \sqrt{2E_e(E+P_Z)}
\end{eqnarray}
where $(E+P_Z)$ is the sum of the energy and $P_Z$ contributions from
the positron and all jets satisfying $P_T^{j}> 15$~GeV and 
pseudorapidity $\eta_{j}<3$ (with the highest $P_T$ jet required to be
outside the FCAL inner ring).
The $\eta_j$ cut removes contributions from the proton remnant.
The constraints $\ptmiss=0$ and $E-P_Z=2E_e$, which are satisfied by
fully contained events, have been assumed in 
arriving at this equation.  When using this mass-reconstruction
method, events with measured
$E-P_Z<40$~GeV were removed to avoid large initial-state QED radiation.

The $M_{CJ}$ method gave, on average, improved resolution 
over the $M_{ej}$ method for narrow LQ MC events.  The improved resolution
occurred at smaller $\cos \theta^{\ast}$
(for $\cos \theta^{\ast} \approx 0$
the mass resolution determined from a Gaussian fit to the reconstructed
mass distribution for $m_{LQ}=200$~GeV was about $1.5$\% in the $M_{CJ}$
method and $3$\% for the $M_{ej}$ method); at the larger 
$\cos \theta^{\ast}$ values where NC DIS events are concentrated, the 
resolutions of the two methods were similar (about $3$\% for $m_{LQ}=200$~GeV).
The $M_{CJ}$ method relies on constraints which do not necessarily apply to a 
resonant state whose properties cannot be anticipated in detail.  
We therefore choose to use the $M_{ej}$ method as our primary search method.  The $M_{CJ}$ method is used in the limit-setting procedure.

\begin{figure}[hbpt]
\begin{center}
\epsfig{figure=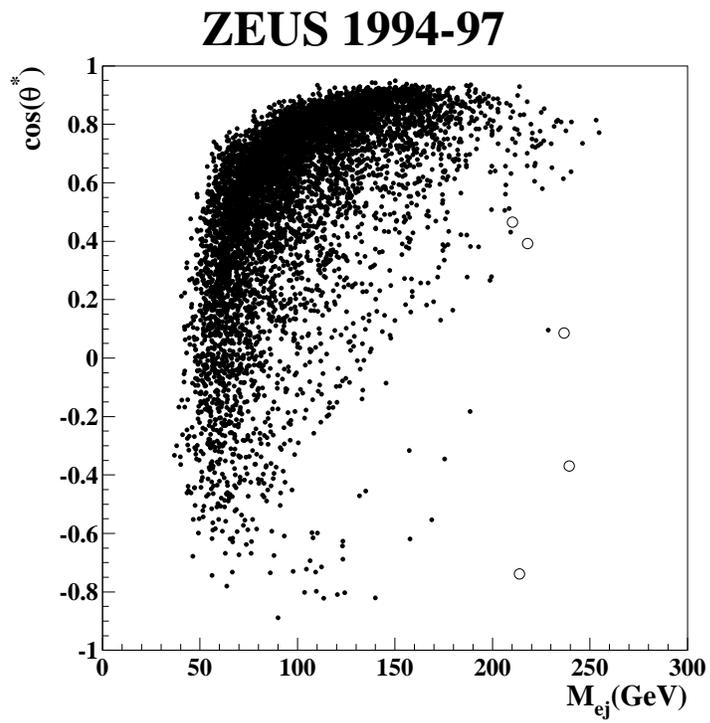,height=10.cm}
\end{center}
\caption{\it The value of $\cos \theta^{\ast}$ versus $M_{ej}$ 
for all 7103 events passing the
event selections.  The events at high $x$ and $y$ described 
in a previous paper
\cite{ref:ZEUShighx} are shown as open circles.}
\label{fig:cost_Mej}
\end{figure}

\section{$M_{ej}$ and $\cos\theta^{\ast}$ Distributions}

The reconstructed values of $M_{ej}$ are plotted 
versus $\cos\theta ^{\ast }$
for the selected events in Fig.~\ref{fig:cost_Mej}.  Most of the events
are concentrated at large $\cos \theta^{\ast}$ and small mass, 
as expected from Standard
Model NC scattering.  
The five events indicated as open circles are from data taken in 1994-96,
with total luminosity $20$~pb$^{-1}$.
They were the subject of a previous publication~\cite{ref:ZEUShighx}.  
In this earlier analysis,
the kinematic variables were reconstructed with the DA method.
The five events also stand out with 
the $M_{ej}$ reconstruction technique. The average value of $M_{ej}$ for
these events is $224$~GeV, or $7$~GeV less than 
the corresponding mass calculated previously
via $M=\sqrt{s\cdot x_{DA}}$, where $x_{DA}$ is the estimator of Bjorken-$x$
calculated with the DA method. 
This mass shift is compatible with expectations based on resolution and
initial state radiation effects.
 With the present 
luminosity of $47.7$~pb$^{-1}$, 7 events are observed in the region 
of $M_{ej}> 200$~GeV and $\cos \theta ^{\ast } < 0.5$, where $5.0$ events
are expected. 

\begin{figure}[hbpt]
\begin{center}
\epsfig{figure=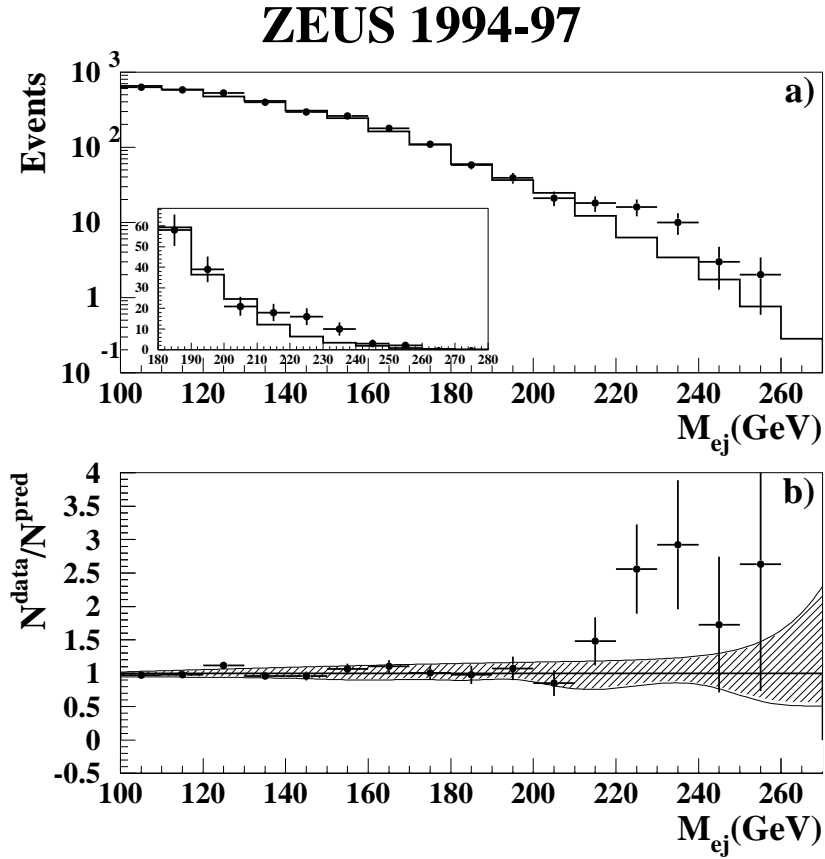,height=12.cm}
\end{center}
\caption{\it a) Comparison of observed events (points) and SM expectations 
(histogram) for 
the reconstructed $e^+$-jet invariant mass. 
The inset shows the region with $M_{ej}>180$%
~GeV on a linear scale. b) The ratio of the 
number of observed
events to the Standard Model expectations.  The shaded band shows the
systematic uncertainty in the predicted number of events. The error
bars on the data points are calculated from the square root of the
number of events in the bin.}
\label{fig:Mej}
\end{figure}

The $M_{ej}$ spectrum for events with $M_{ej}>100$~GeV is shown 
in Fig.~\ref{fig:Mej}a on a logarithmic scale. The
high-mass part of the spectrum is shown on a linear scale in the inset. 
The predicted number of events ($N^{pred}$) from NC processes is shown as 
the histogram.  The ratio of the measured mass spectrum to the expectation 
is shown in Fig.~\ref{fig:Mej}b. The shaded band indicates
the systematic uncertainty on the expectations.

\subsection{Systematic Uncertainties}
\label{sec:systematics}

The uncertainty
on $N^{pred}$ varies with mass from $7$\% at 100 GeV 
up to 30\% at 250 GeV.  The most
important uncertainties are on the energy scale and 
the jet position.  The NC DIS cross section given in
Eq.~\ref{eq:DIS} (neglecting $F_L$) can be rewritten in terms of the $e^+q$ invariant
mass, $M$, and the polar angle of the outgoing struck quark
in the laboratory frame, $\gamma$:
\begin{equation}
\label{eq:Mgamma}
\frac{d^2 \sigma(e^+p)}{dMd\gamma} = \frac{32\pi \alpha^2E_e^2 \sin\gamma}
{M^5(1-\cos\gamma)^2}
\left[Y_+ {\it F_2} - Y_-x{\it F_3} \right]\;\; .
\end{equation}
The mass dependence is very steep.  In addition to the
explicit $M^{-5}$ dependence, there is also a strong suppression of
high masses implicit in the structure functions.  
An incorrect energy scale will produce a shift in the mass spectrum 
and potentially a significant error in the number of expected events 
at a given mass.  
The dependence on the quark angle is also steep, approximately $\gamma^{-3}$
at small $\gamma$.  The number of events passing the jet fiducial
cut is therefore strongly dependent on the accuracy of the
jet position reconstruction.
 The jet fiducial-volume cut requires the
highest-$P_T$ jet to point outside the inner ring of FCAL towers.  Many
distributions from data and MC were compared to search
for possible systematic biases.

The dominant sources of uncertainty are
itemized below in order of decreasing importance:

\begin{enumerate}
\item Knowledge of the calorimeter energy scales:

  The scale uncertainties discussed in Sect.~\ref{sec:conditions} are
$1$\% for BCAL positrons and $2$\% for hadrons, leading to an 
uncertainty of
$5 (18)$\% in $N^{pred}$ at $M_{ej}=100 (210)$~GeV.

\item Uncertainties in the simulation of the hadronic energy flow,
including simulation of the proton remnant, the energy flow
between the struck quark and proton remnant, and possible
detector effects in the innermost calorimeter towers:

Many distributions of data and MC were compared and
no important systematic differences were found.
Figure~\ref{fig:eflow} shows the fraction of the jet energy 
in the inner ring of FCAL towers associated with the
highest $P_T$ jet as a function of $\eta_{j}$.  This is shown
for all events in Fig.~\ref{fig:eflow}a, as well as for those with 
$M_{ej}>210$~GeV in Fig.~\ref{fig:eflow}b.
For the
highest $\eta_{j}$ values considered, this ratio is about $20$\%.
The energy located in the
innermost towers of the FCAL and not associated with the highest $P_T$
jet is shown in Fig.~\ref{fig:eflow}c,d, and compared to the MC
simulation.  No large differences are seen between data and MC
(the lowest $\eta$ bin in Fig~\ref{fig:eflow}d
 contains only five data events).
The innermost towers of the FCAL have a larger 
uncertainty in the energy scale than the rest of the 
FCAL owing to their slightly different
construction and proximity to the beam. The
energy in these cells has been varied by $\pm 10$\%. 
As a test of the simulation
of the forward energy flow, the ARIADNE MC has been used instead of
the LEPTO MC.  These tests yielded variations in $N^{pred}$ of 
13\% at $M_{ej}=210$~GeV.

\begin{figure}[hbpt]
\begin{center}
\epsfig{figure=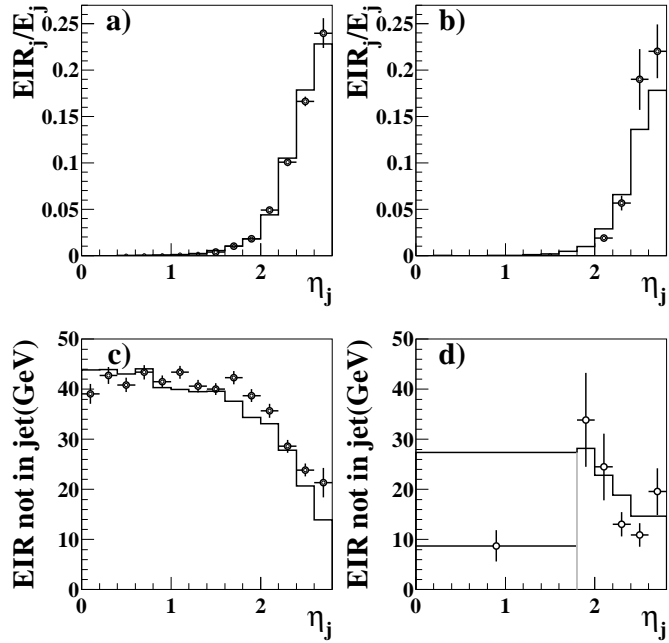,height=10.cm}
\end{center}
\caption{\it The ratio of the jet energy in the innermost towers
of the FCAL, EIR$_j$, to the total jet energy, $E_j$,
as a function of $\eta$ of the jet for a)
the full sample, and b) for those events with $M_{ej}>210$~GeV.
The energy deposited in the
innermost FCAL towers, excluding that associated with the highest $P_T$
jet, is shown in c) for the full sample, and in d) for those
events with $M_{ej}>210$~GeV.  The data are shown as points, while the
NC Monte Carlo predictions are shown as a histogram.}
\label{fig:eflow}
\end{figure}

\item Uncertainty in the parton density functions:

The parton density functions were
estimated as in~\cite{ref:highQ2}, and led to an uncertainty of 5\%
in $N^{pred}$ at $M_{ej}=210$~GeV.

\item Uncertainties in the acceptance: 

The alignment of the FCAL was determined
to better than $5$~mm, and various jet position reconstruction
algorithms were compared. These studies yielded an
uncertainty of 2\% in $N^{pred}$.

\item Uncertainties in the energy resolution functions. 

 These were
studied by comparing tracking information with calorimeter information
for individual events,
as well as by comparing different reconstruction methods.  The
MC energies were smeared by additional amounts to represent these
uncertainties, leading  to a variation
of less than $5$\% in $N^{pred}$.  

\end{enumerate}

Other uncertainties include
positron finding efficiency, 
luminosity determination, vertex simulation, multijet production rates,
and hadronization simulation.  
These were found to be 
small in comparison to the items listed above. 
The overall systematic uncertainty was obtained by
summing the contributions from all these sources in quadrature.

\subsection{Discussion}

\begin{figure}[tbph]
\begin{center}
\epsfig{figure=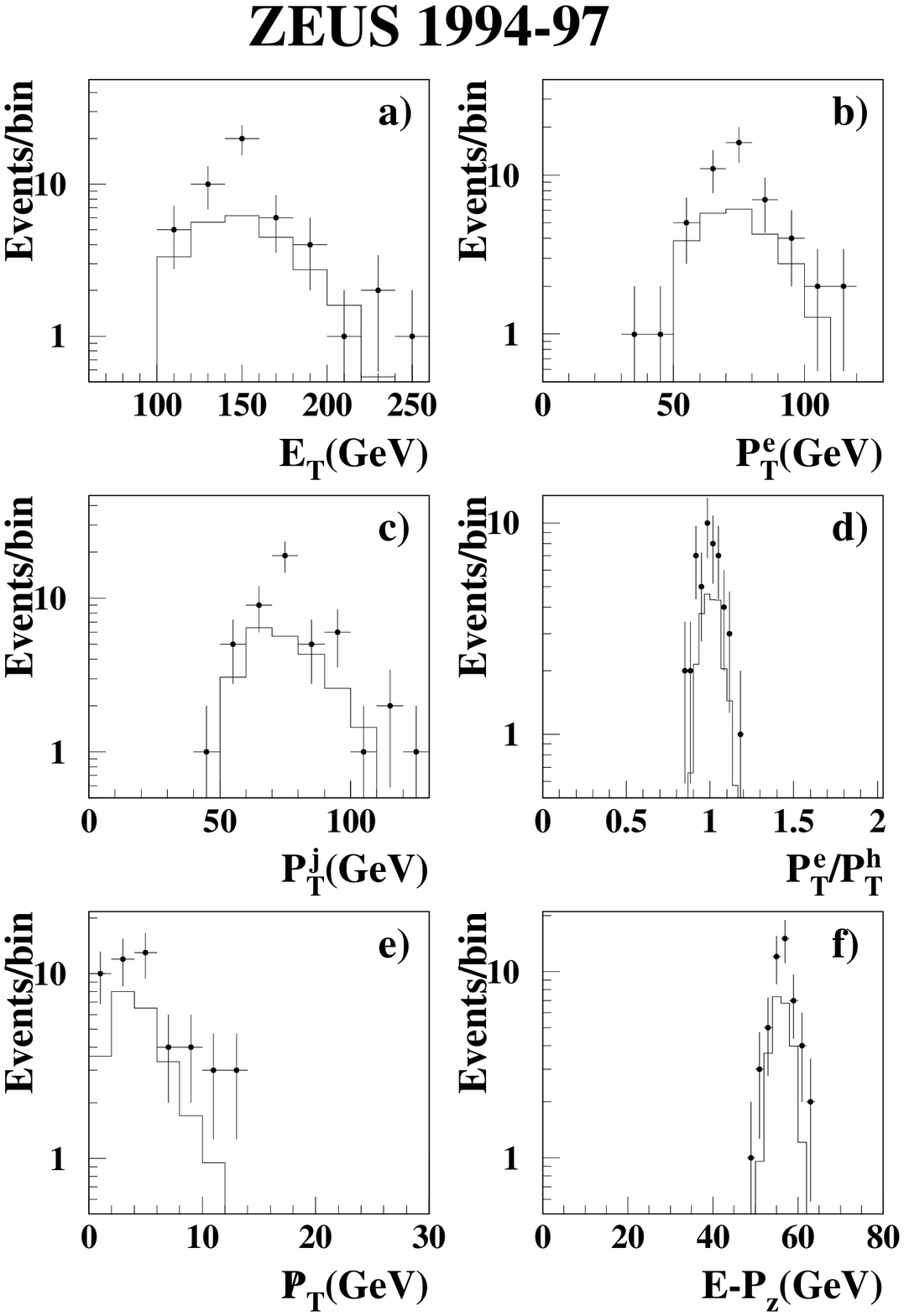,height=16.cm}
\end{center}
\caption{\it Comparison of data (points)
with Standard Model expectations (histograms) for
selected distributions and requiring
$M_{ej}>210$~GeV: a) total event transverse energy, $E_{T}$; b) \
positron transverse momentum, $P_{T}^{e}$;
 c) jet transverse momentum, $P_{T}^{j}$%
; d) the ratio of the positron to hadron transverse momenta, 
$P_{T}^{e}/P_{T}^{h}$, (e) the net (or missing) transverse momentum, \ptmiss, 
and (f) the longitudinal momentum variable, $E-P_{Z}$, for the event.}
\label{fig:compare_210}
\end{figure}

The data in Fig.~\ref{fig:Mej} are in good agreement with the SM expectations
up to $M_{ej}\approx 210$~GeV.  Some excess is seen at higher
masses.
 For $M_{ej}>210$~GeV, $49$ events were observed in the data,
while $24.7 \pm 5.6$ events are expected. A careful study
of individual  events in this mass region
uncovered no signs of reconstruction errors.
Rather, the events always contain clear examples 
of a high-energy positron (typically $70$~GeV) near $90^{\circ}$ and a 
high-energy jet (typically $400$~GeV)
in the forward direction (2 events have a second jet, in accord with NC
DIS Monte Carlo expectations). The distributions shown in Fig.~\ref{fig:compare}
for all the data are restricted to the
events with $M_{ej}>210$ GeV in Fig.~\ref{fig:compare_210}. 
Whereas the shapes of the distributions
are similar, the data lie systematically above the MC, which is normalized to
the integrated luminosity.

\begin{figure}[hbpt]
\begin{center}
\epsfig{figure=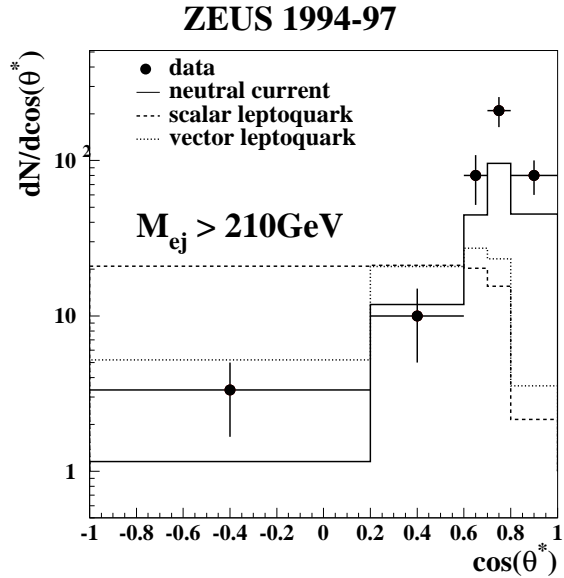,height=8.cm}
\end{center}
\caption{\it Comparison of data (points) and NC Monte Carlo expectations 
(solid histogram) for the $\cos \theta ^{\ast }$ distribution for events 
with $M_{ej}>210$~GeV. 
The predictions for narrow scalar and vector resonant leptoquarks 
are shown with arbitrary normalization for comparison.}
\label{fig:cost}
\end{figure}

The events with large $M_{ej}$ have characteristics
similar on average to NC DIS events.
In particular, the $\cos \theta
^{\ast }$ projection of the events with $M_{ej}>210$~GeV is shown in Fig.~%
\ref{fig:cost} and compared to the MC expectations for neutral current
DIS (solid histogram).  
The expectations for narrow s-channel scalar and vector
LQ production are also shown for comparison.  For $F=0$ LQs with 
$\lambda<1$,
the u-channel and interference terms would not significantly affect these
expectations.
The shape of the data and NC MC $\cos \theta^{\ast}$
distributions are qualitatively similar, peaking at high
values of $\cos \theta ^{\ast }$.

In summary, there is some excess of events with $M_{ej}>210$~GeV above the 
Standard Model
predictions.  The probability of observing such an excess depends strongly
on possible systematic biases.  The most important of these are
biases in the energy
scales.  As a test, many MC experiments were generated where  the jet 
energy scale was shifted by +2\% and the electron energy scale by +1\%.  
A window of width $3\sigma(M_{ej})$, where $\sigma(M_{ej})$ is the mass
resolution at mass $M_{ej}$, was moved over the accessible mass
range.  For each simulated experiment, the number of observed events within
the mass window was compared with the nominal 
expectations as a function of $M_{ej}$,
seeking the excess which gave the largest statistical significance.  
The same procedure was applied to the data.
As a result, it was found that 5\% of the simulated experiments would observe,
somewhere in the mass spectrum, an excess of statistical
significance at least as large as the one found in the data.
The excess is therefore not statistically compelling.  Furthermore,
the events have the characteristics of neutral current scattering.  Limits
are therefore set on the production of narrow scalar or vector
states, as discussed below.

\section{Limits on Narrow Scalar and Vector States}
\label{sec:limits}

Limits are set on the production cross section times branching ratio
into positron+jet,
$\sigma B$, for a narrow scalar or vector state. For definiteness,
limits on coupling strength versus mass for $F=0$ leptoquarks
are presented, as well as limits on $\lambda \sqrt{B}$ versus mass for
scalar states coupling to $u$ or $d$ quarks, such as $\rpvio$ squarks.
The limits are extracted for $\lambda \leq 1$, allowing the
use of the
narrow-width approximation assumed in Eq.~\ref{eq:NWA}.

\begin{figure}[hbpt]
\begin{center}
\epsfig{figure=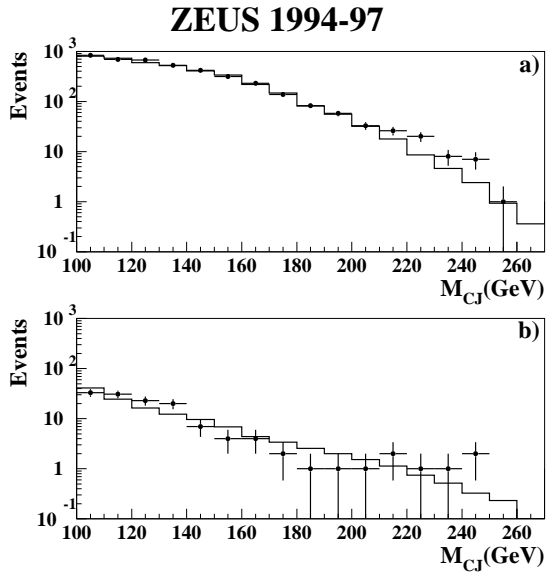,height=8.cm}
\end{center}
\caption{\it The reconstructed mass spectrum using the $M_{CJ}$ method
for data (points) and SM expectations (histogram):
a) shows the spectrum for events satisfying all cuts, while
b) shows the mass spectrum after the cut
($\cos\theta^{\ast} < \cos \theta^{\ast}_{cut}$) for the scalar LQ search 
has been applied.}
\label{fig:MCJ}
\end{figure}

The $M_{CJ}$ mass
reconstruction method was used to set limits as described in 
Sect.~\ref{sec:Reconstruction}.  The positron fiducial cuts were removed 
since this method is less sensitive to the
positron-energy measurement, while the cut $E-P_Z>40$~GeV was applied
to reduce radiative effects.   The
mass spectrum reconstructed with this technique is shown in 
Fig.~\ref{fig:MCJ}a.  In total, 8026 events passed all selection cuts while
7863 events are predicted by the MC.

\begin{figure}[hbpt]
\begin{center}
\epsfig{figure=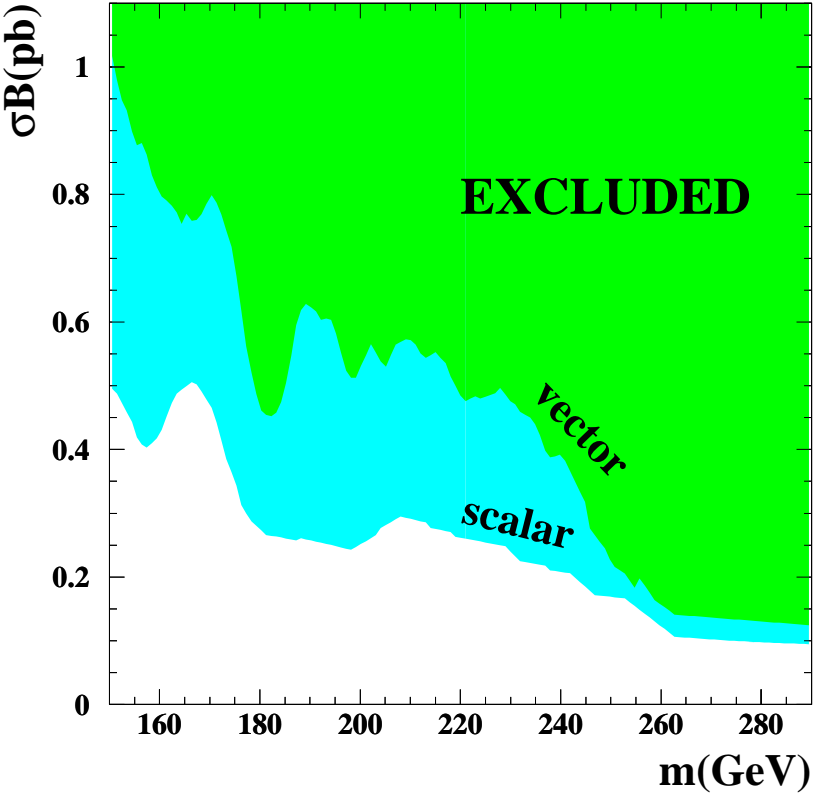,height=12.cm}
\end{center}
\caption{\it Limits on the production cross section times branching ratio 
for decay into $e^+$-jet(s) for a scalar or vector state, as a function
of the mass of the state.  The shaded regions
are excluded.}
\label{fig:CSL_Mejet}
\end{figure}

The leptoquark MC described in Sect.~\ref{sec:MC} 
was used to determine the event selection efficiency
and the acceptance of the fiducial cuts, as well as to estimate
the mass resolution. 
This MC and the NC\ background simulation were used to
calculate an optimal bin width, $\Delta M_{CJ}$, for each $M_{CJ}$, 
and
optimal  $\cos \theta^{\ast }$ range, $\cos \theta^{\ast}<
\cos \theta^{\ast}_{cut}$, 
to obtain on average the best limits on LQ couplings.  
The bin widths were typically 20 GeV.
The values of $\cos \theta _{cut}^{\ast }$ for setting limits range
from $0.5$ to $0.9$ for vector leptoquarks with masses between $150-290$%
~GeV, and from $0.1$ to $0.9$ for scalar leptoquarks in the same mass
range.  The mass spectrum after applying the optimal $\cos \theta^{\ast}$
cut for the scalar search is shown in Fig.~\ref{fig:MCJ}b.
No significant deviations from expectations are seen after applying this
cut.

The 95\% confidence level (CL) limits on $\sigma B$ were
obtained directly from the observed number of data events with $\cos \theta
^{\ast } < \cos \theta _{cut}^{\ast }$ \ in the particular mass 
window~\cite{ref:PDG}.  The procedure described in ~\cite{ref:PDG} was
extended to include the systematic uncertainties in the numbers of
predicted events.  This was found to have negligible effect on the limits.
The limits for a narrow scalar
or vector state are shown in Fig.~\ref{fig:CSL_Mejet}.  These limits
lie between $1$ and $0.1$~pb as the mass increases from $150$ to
$290$~GeV.

\begin{figure}[hbpt]
\begin{center}
\epsfig{figure=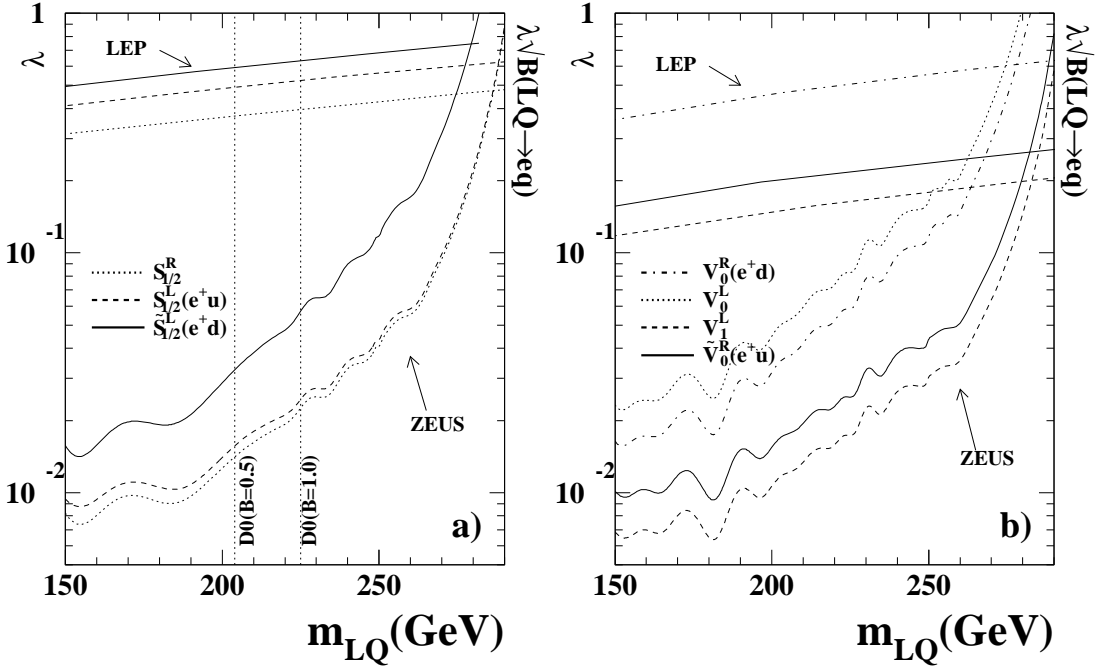,height=10.cm}
\end{center}
\caption{\it Coupling limits as a function of leptoquark
mass for $F=0$ leptoquarks.  The results from this analysis 
are compared to representative limits from LEP~\cite{ref:OPAL} and the
Tevatron~\cite{ref:D0}. The areas above the ZEUS and LEP curves 
are excluded, while the area to the left of the Tevatron line is excluded
for scalar leptoquarks with the indicated branching ratio to $e$+jet. The limits
on scalars are shown in a) while the limits on vectors are shown in b).}
\label{fig:LQlimits}
\end{figure}

The 95\% CL exclusion limits for different species of LQ are given in the
coupling versus mass plane in Fig.~\ref{fig:LQlimits}. 
The limits exclude leptoquarks with coupling strength
$\lambda=\sqrt{4\pi\alpha}\approx 0.3$ for masses up to
280 GeV for specific types of F=0 leptoquarks. 
The H1 collaboration has recently
published similar limits~\cite{ref:H1_limits}.
In Fig.~\ref{fig:LQlimits}, the ZEUS results are compared to 
recent limits from OPAL.  
At LEP~\cite{ref:L3,ref:OPAL,ref:ALEPH}, sensitivity to a high-mass LQ 
arises from effects of
virtual LQ exchange on the 
hadronic cross section.  The HERA and LEP 
limits are complementary to Tevatron limits~\cite{ref:CDF,ref:D0}, 
which are independent of the coupling 
$\lambda_{L,R}$. The limits by D0 (CDF)
extend up to $225 \; (213)$~GeV for a scalar LQ
with 100\% branching ratio to $eq$. 
The D0 limits are shown as 
vertical lines in Fig.~\ref{fig:LQlimits}.  The Tevatron limits for vector
LQs are model dependent~\cite{ref:Bluemlein}, but are expected to be
considerably higher than for scalar LQs.  

The ZEUS limits presented in Fig.~\ref{fig:LQlimits} can also be applied to any narrow
state which couples to a positron and a $u$ or $d$ quark and with unknown branching
ratio to $e^+$-jet(s).  These states correspond to the leptoquark types as labelled
in the figure.  For these states, the limits are on the quantity
$\lambda \sqrt{B}$. Examples of scalar states for
which these limits apply are $\rpvio$-squarks (e.g.,the limit on the
$\tilde{S}_{1/2}^L(e^+d)$ LQ can be read as a limit on the
$\lambda_{1j1}'$ R-parity-violating coupling).

\section{Conclusion}

Data from $47.7$~pb$^{-1}$ of $e^+ p$ collisions at a center-of-mass energy 
of 300~GeV have been used to search
for a resonance decaying into $e^+$-jet.
The invariant mass of the $e^+$-jet pair was calculated directly from
the measured energies and angles of the positron and jet.  This
approach makes no assumptions
about the production mechanism of such a state.

The observed mass spectrum is in good agreement with Standard Model 
expectations up to $e^+$-jet masses of about $210$~GeV.   Above this
mass, some excess is seen. The angular distribution of these events 
is typical of high-$Q^2$ neutral current events and does not give
convincing evidence for the presence of a narrow scalar or vector state.  
By applying restrictions on the decay angle to optimize sensitivity to a
narrow state in the presence of NC background, 
limits have been derived 
on the cross section times decay branching fraction
for a scalar or vector state decaying into positron and jet(s). 
 These limits can be interpreted, for example,
as limits on leptoquark or R-parity-violating squark production.
Limits on the production of leptoquarks and squarks are presented in the
coupling strength versus mass plane. At a coupling strength
$\lambda=0.3$, new states are ruled out at 95\% confidence level
for masses between $150$ and  $280$~GeV.

\section*{Acknowledgements}
We thank the DESY Directorate for their strong support and encouragement,
and the HERA machine group for their diligent efforts.
We are grateful for the  support of the DESY computing
and network services. The design, construction and installation
of the ZEUS detector have been made possible by the ingenuity and effort
of many people from DESY and home institutes who are not listed
as authors. It is also a pleasure to thank
W. Buchm\"uller, R. R\"uckl and M. Spira for useful discussions.


\begin{thebibliography}{99}

\bibitem{ref:H1highx} H1 Collab., C. Adloff et al., 
Z. Phys. C74 191 (1997); 
H1 Collab., DESY 99-107, Accepted by Eur. Phys. J. C

\bibitem{ref:H1_limits} H1 Collab., C. Adloff et al.,
Eur. Phys. J. C11 447 (1999) and erratum ibid

\bibitem{ref:ZEUShighx} ZEUS Collab., J. Breitweg et al., 
Z. Phys. C74 207 (1997)

\bibitem{ref:highQ2} ZEUS Collab., J. Breitweg et al., 
Eur. Phys. J. C11 427 (1999) 

\bibitem{ref:dokshitzer} Yu. L.~Dokshitzer, Proc. of the 5th International
Workshop on Deep Inelastic Scattering and QCD, Chicago, IL, USA,
Eds. J. Repond and  D. Krakauer, American Institute of Physics (1997)


\bibitem{ref:lq} W. Buchm\"{u}ller, R. R\"{u}ckl and D. Wyler, 
Phys. Lett. B191 442 (1987); erratum Phys. Lett. B448 320 (1999)

\bibitem{ref:Rpsusy} J. Butterworth and H. Dreiner, Nucl. Phys.
B397 3 (1993), and references therein

\bibitem{ref:SF} G. Ingelman and R. R\"uckl, Phys. Lett. B201 369 (1988)

\bibitem{ref:Davidson} S. Davidson, D. Bailey and B. Campbell, Z. Phys. 
C61 613 (1994)

\bibitem{ref:LG} B. Schrempp, Proc. of the Workshop
``Physics at HERA'', vol. 1, Eds. W. Buchm\"uller and G. Ingelman, DESY 
1034 (1991)

\bibitem{ref:DURHAM}
T.~Matsushita, E.~Perez and R.~R\"uckl, Proc. of the 3rd UK Phenomenology
Workshop on HERA Physics, Durham, England (1998), J. Phys. G25 1418 (1999) 

\bibitem{ref:lqQCD} T. Plehn, H. Spiesberger, M. Spira and P. M. Zerwas, 
Z. Phys. C74 611 (1997); \\
Z. Kunszt and W. J. Stirling, Z. Phys. C75 453 (1997)



%
\bibitem{ref:Hewett_SUSY} J. L. Hewett, `Single squark production at HERA via 
R parity violating interactions', Proc. of the 1990 DPF Summer Study on 
High-Energy Physics, Snowmass, Colorado, USA, Eds. E. Berger, 
World Scientific (1992)

\bibitem{ref:ZEUS} ZEUS Collab., ``The ZEUS Detector'', 
Status Report 1993, DESY 1993


\bibitem{ref:DA} S. Bentvelsen, J. Engelen and 
P. Kooijman, Proc. of the Workshop
``Physics at HERA'', vol. 1, Eds. W. Buchm\"uller and G. Ingelman, DESY 
23 (1991) ; K. C. Hoeger, ibid. 43


\bibitem{ref:lumi} J. Andruszk\'ow et al., DESY 92-066 (1992);
ZEUS Collab., M. Derrick et al., Z. Phys. C63 391 (1994) 
 
\bibitem{ref:kT1} S. Catani, Yu. L. Dokshitzer, M. H. Seymour and 
B. R. Webber, Nucl. Phys. B406 187 (1993) 

\bibitem{ref:kT2} S. D. Ellis and D. E. Soper, Phys. Rev. D48 3160 (1993)

\bibitem{ref:HERACLES} HERACLES 4.4: A. Kwiatkowski, H. Spiesberger and
H.-J. M\"ohring, Comput. Phys. Commun. 69 155 (1992) 

\bibitem{ref:DJANGO} DJANGO 1: K. Charchula, G. A. Schuler and H. Spiesberger,
Comput. Phys. Commun. 81 381 (1994)

\bibitem{ref:MEPS} LEPTO 6.1: G. Ingelman, A. Edin and J. Rathsman, Comput. 
Phys. Commun. 101 108 (1997)

\bibitem{ref:ARIADNE} ARIADNE 4: L. L\"onnblad, Comput. Phys. Commun. 71 
15 (1992)

\bibitem{ref:CTEQ4} CTEQ Collab., H. L. Lai et al., 
Phys. Rev. D55 1280 (1997)


\bibitem{ref:PYTHIA} PYTHIA 6.1, C. Friberg, E. Norrbin and T. Sj\"ostrand,
Phys. Lett. B403 329 (1997) 

\bibitem{ref:GEANT} GEANT 3.13: R. Brun et al., CERN-DD/EE/84-1 (1987)


\bibitem{ref:PDG} 
Particle Data Group, K. Hikasa et al., Phys. Rev. D45 S1 (1992)

\bibitem{ref:L3}
L3 Collab., M.~Acciarri et al., Phys. Lett. B433
163 (1998)

\bibitem{ref:OPAL}
OPAL Collab., G.~Abbiendi et al.,
Eur. Phys. J. C6 1 (1999)

\bibitem{ref:ALEPH}
ALEPH Collab., R.~Barate et al.,
{\rm hep-ex/9904011}

\bibitem{ref:CDF} CDF Collab., F. Abe et al., 
Phys. Rev. Lett. 79 4327 (1997) 

\bibitem{ref:D0} D0 Collab., B. Abbott et al., Phys. Rev. Lett. 
80 2051 (1998)

\bibitem{ref:Bluemlein} J. Bl\"umlein, E. Boos and A. Kryukov, Z. Phys. 
C76 137 (1997)

\end{thebibliography}
\end{document}